# Domain-Agnostic Causal-Aware Audio Transformer for Infant Cry Classification

Geofrey Owino
School of Computing and Engineering Sciences
Strathmore University
Nairobi, Kenya
geoffreyowino@strathmore.edu

Bernard Shibwabo Kasamani
School of Computing and Engineering Sciences
Strathmore University
Nairobi, Kenya
0000-0002-0827-9857

Ahmed M. Abdelmoniem
School of Electronic and Computer Engineering Science
Queen Mary University of London
London, UK
ahmed.sayed@qmul.ac.uk

Edem Wornyo
Google Research
Google LLC New York
New York, USA
wornyo@google.com

*Abstract*— Accurate and interpretable classification of infant cry paralinguistics is vital for early diagnosis of neonatal distress and clinical decision support. However, most existing deep learning approaches rely heavily on correlation-driven signal representations, making them vulnerable to acoustic perturbations and domain shifts across recording environments. In this work, we present DACH-TIC, a novel Domain-Agnostic Causal-Aware Hierarchical Audio Transformer that integrates causal reasoning, multi-task modeling, and adversarial domain generalization for robust infant cry classification. DACH-TIC introduces a structured transformer backbone composed of local token-level and global semantic encoders, augmented by causal attention masking and a controlled perturbation training approach that simulates counterfactual acoustic perturbations. A domain adversarial head further enables invariance across recording environments, while multi-task supervision improves representation robustness by jointly optimizing cry type, distress intensity, and causal relevance. We evaluate DACH-TIC on cry recordings from the Baby Chillanto and Donate-a-Cry datasets, with ESC-50 noise overlays for domain augmentation. Compared to existing state-of-the-art models such as Hierarchical Token-Semantic Audio Transformer (HTS-AT) and Squeeze-and-Excitation Residual Network Transformer (SE-ResNet Transformer), DACH-TIC achieves significant improvements in accuracy (↑2.6%), macroF1 (↑2.2), and causal fidelity metrics. It also generalizes well to unseen environments with minimal performance degradation (domain gap: 2.4%). These results establish DACH-TIC as a causally grounded, domain-resilient model for real-world deployment in neonatal acoustic monitoring systems.

*Keywords*— *Domain-agnostic, Causal-aware, Transformer, Paralinguistics, Domain-adversarial, Pseudo-interventional.*

## I. Introduction

Crying remains the primary communication modality in infants, often signaling critical physiological or emotional states such as hunger, pain, fatigue, or discomfort[1]. Classifying these cries with high precision enables timely interventions, particularly in neonatal intensive care units (NICUs) and low resource environments[2, 3]. Ji et al. have shown that early paralinguistic analysis of infant cries holds predictive power for neurological and developmental outcomes[4, 5]. Recent evidence also shows that neonatal cry acoustics are associated with later neurodevelopment[6]. A multicenter cohort of very preterm infants found cry features at NICU discharge linked to cognitive, language, and behavioral outcomes at 2 years[7]. Lawford et al. in their review likewise supports that cry acoustics are markers of early neurological dysfunction[6].

Although recent advancements in deep learning have yielded notable gains in infant cry classification, most existing models, including those proposed by Teeravajanadet et al. and Maghfira et al., are based on convolutional or recurrent architectures [8, 9]. These models often learn spurious correlations in the input data, which degrades their generalization performance in real-world settings characterized by domain shifts, such as noise variation or microphone differences[10].

Transformer-based architectures like Hierarchical Token Semantic Audio Transformer (HTS-AT), proposed by Chen et al., have achieved strong performance by modeling temporal token dependencies through attention mechanisms [11]. However, these models are still vulnerable to overfitting to non-causal acoustic artifacts, especially when trained without explicit structural priors. Baevski et al. further demonstrated the limits of unsupervised speech transformers in generalizing across acoustic domains [12].

According to Jiao et al., domain-specific noise and environmental conditions can introduce confounding artifacts that affect the learned representation [13]. Lin et al. argued that Bayesian risk minimization is a powerful alternative to empirical risk minimization in such non-independent and identically distributed conditions [14]. Similarly, Scholkopf et al. emphasized that models aligned with causal principles are more likely to generalize under interventions and distribution shifts [15].

Existing infant-cry models and audio transformer approaches optimize empirical risk with bidirectional, unconstrained attention and lack mechanisms for temporal causality, intervention-level robustness, and joint domain invariance[8, 9, 11]. Single-stage encoders mix local transients with global semantics, and most explanations are post hoc rather than supervised[16, 17]. DACH-TIC addresses these limitations with a hierarchical token-to-semantic encoder and a causal attention mask that restricts attention to past evidence. It adds consistency under controlled perturbations of non-causal regions and a GRL-based domain head for environment invariance. Multi-task heads for cry type, distress intensity, and token-level relevance are trained with a composite loss to support robust and interpretable performance.

This work was supported in part by the Google Research Grant.

*A. Objectives*

This work was driven by the following objectives:

i. To develop a causal-aware transformer model capable of prioritizing semantically and developmentally relevant acoustic cues using attention masking and pseudo intervention training.

ii. To promote robustness across domains by employing adversarial learning mechanisms that minimize the influence of environment-specific acoustic signatures.

iii. To optimize the model across multiple complementary tasks: cry type classification, distress intensity regression, and causal salience estimation.

iv. To evaluate the model's effectiveness on real-world infant cry datasets with realistic noise overlays, benchmarking against state-of-the-art models under classification, robustness, and domain transfer metrics.

*B. Contributions*

The key contributions of this paper are as follows:

i. Domain-Agnostic Causal-Aware Hierarchical Audio Transformer (DACH-TIC) with a two-stage token-to-semantic encoder and causal attention masking.

ii. Controlled perturbation training that enforces prediction consistency under non-causal spectrogram edits to improve robustness.

iii. Domain-adversarial head with gradient reversal to learn environment-invariant features.

iv. Multi-task supervision with heads for cry type, distress intensity, and token-level relevance optimized via a composite loss.

v. Empirical gains over HTS-AT[11] and SE-ResNet-Transformer[18] on Baby Chillanto and Donate-a-Cry with ESC-50 overlays, achieving 97.6% accuracy, 0.941 macro-F1, 0.98 AUC, and a 2.4% domain-gap.

## II. RELATED WORK

The study of infant cry classification intersects multiple research domains, including practical audio analysis, causal representation learning, and domain-generalizable architectures. This section critically reviews foundational and contemporary efforts across these fronts, highlighting key gaps that motivate the proposed DACH-TIC architecture.

*A. Deep Learning for Infant Cry Paralinguistics*

The initial models developed for infant cry recognition predominantly relied on handcrafted features such as fundamental frequency, harmonic ratios, or Mel-frequency Cepstral Coefficients (MFCCs) to distinguish cry types [19, 20]. These features, while interpretable, lack robustness in non-stationary or noisy conditions. The shift toward data-driven learning began with Convolutional Neural Networks (CNNs) trained on spectrograms[21]. Teeravajanadet et al. (2019) demonstrated that CNN-based models can achieve moderate accuracy in distinguishing hunger and pain cries however, their architectures were sensitive to background noise and lacked temporal modeling capabilities [8, 21].

To improve temporal context modeling, Maghfira et al. proposed a hybrid CNN–RNN architecture for infant cry classification, demonstrating improved performance over conventional CNNs due to better sequence modeling[9]. However, despite improved performance with CRNN the model is correlation-driven and lacked robustness to distributional shifts across recording environments [9].

More recently, transformer-based models have emerged as the dominant paradigm in audio classification. The HTS-AT model introduced by Chen et al. employed a hierarchical transformer structure to capture both token-level and semantic level information from log-Mel spectrograms [11]. Complementarily, the Audio Spectrogram Transformer (AST) used by Gong et al. is a convolution-free, patch-based audio transformer pretrained on AudioSet that models global time–frequency dependencies with bidirectional attention[22]. While HTS-AT and AST achieved state-of-the-art performance on general audio tasks, they lacked both causal interpretability and robustness to environment-specific noise, two aspects critical for infant cry analysis in real-world deployment[11, 22].

*B. Causal Representation Learning in Audio Modeling*

The integration of causal principles into machine learning has gained attention as a means to achieve out-of-distribution generalization and interpretable decision-making. Scholkopf et al. argued that models grounded in causal assumptions are more likely to remain stable under interventions or environmental shifts [15]. In the audio domain, recent work emphasize that learned representations should separate semantically relevant causal features from spurious artifacts like ambient noise or pitch variations [13, 23]. However, these insights have not been operationalized in infant cry modeling, where cries may exhibit natural variability tied to both biological and contextual factors.

Recent techniques for causal-aware learning include temporal masking guided by attention priors [24], counterfactual consistency constraints [25], and simulation of acoustic interventions via targeted perturbations.[26]. Yet, these have rarely been embedded into transformer-based architectures for health-oriented audio classification. DACH-TIC explicitly addresses this by combining causal masking, perturbation-based pseudo-intervention training, and per-token causal salience estimation offering a unified and operational causal modeling strategy within a deep transformer backbone.

*C. Domain Adaptation and Generalization in Acoustic Systems*

Infant cry classification is particularly susceptible to domain shifts caused by differences in microphone type, background acoustic profile (e.g., home vs. NICU), or recording conditions. Lin et al introduced the concept of Bayesian Invariant Risk Minimization (BIRM) as a theoretical strategy for learning features that are stable across multiple domains [14]. While theoretically powerful, BIRM has seen limited practical implementation in cry classification systems.

A more common alternative involves adversarial training using domain confusion objectives. The Domain-Adversarial Neural Network (DANN) proposed by Ganin et al introduced the use of a Gradient Reversal Layer (GRL) to encourage feature encoders to remove domain-identifying information [27]. This approach has been applied in environmental sound classification but remains underutilized in the context of neonatal audio.

Baevski et al. introduced wav2vec 2.0 as a self-supervised model capable of learning universal speech representations [12]. However, such models focus on global signal reconstruction and often neglect causal alignment or domain-specific adaptation. DACH-TIC advances the field by integrating GRL based domain adversarial training with a transformer encoder tuned for infant vocalization analysis, thereby reducing domain generalization gaps and improving transferability.

*D. Summary and Research Gap*

Across the reviewed literature, three core limitations emerge: (i) most models rely on correlation-based learning, ignoring the causal mechanisms underlying cry generation; (ii) generalization to unseen environments remains poor due to the absence of domain-robust design principles; and (iii) multi-task learning architectures that jointly optimize cry classification with auxiliary tasks like intensity estimation or causal relevance have not been fully explored. DACH-TIC is positioned to address all three limitations through a unified architecture that incorporates hierarchical temporal modeling, causal-aware attention, domain-adversarial optimization, and multi-task outputs.

## III. METHODOLOGY

*A. Dataset Diversity And Potential Biases*

We used Baby Chillanto and Donate-a-Cry with multiple infants across genders and early infancy, covering five cry categories (belly pain, burping, discomfort, hunger, tired). Recordings span NICU, home, and outdoor conditions with varied microphones and rooms. For augmentation, ESC-50 dataset[28] adds indoor/domestic and outdoor/urban backgrounds as well as human non-speech and natural sounds at multiple SNRs.

Potential biases could be attributed to an uneven mix of clean and noisy recordings across sources, which can couple environmental conditions to labels and create spurious associations. They could also arise from overlapping acoustic cues between belly pain and burping, which reduces class separability. Age-related variation in early infancy could further shift pitch and prosody and alter feature distributions.

*B. Input Representation and Domain-Augmented Preprocessing*

Each infant's cry waveform was resampled to 16 kHz to standardize the temporal resolution across datasets. A log-Mel spectrogram representation was then computed from the waveform $x(t) \in \mathbb{R}^T$ by applying the short-time Fourier transform (STFT), followed by Mel-scale filtering and logarithmic compression. The resulting spectrogram $X \in \mathbb{R}^{T' \times F}$ was defined element-wise as:

$$X(t', f) = \log\left(\sum_{k=1}^{K} M_{f,k} \cdot \left|\text{STFT}(x(t))_k\right|^2 + \epsilon\right) \quad (1)$$

where $M_{f,k}$ denoted the Mel Filter-bank matrix, $K$ represented the number of frequency bins, and $\epsilon$ was a small constant for numerical stability. This representation preserved the time-frequency energy distribution of the cry signal in a perceptually relevant space.

Additive environmental noise was incorporated to account for domain variability and prepare the model for real-world deployment scenarios. Specifically, background signals $x_n(t)$ were sampled from the ESC-50 dataset (Kaggle) and mixed with the cry waveform using a scalar mixing factor $\alpha \in [0,1]$, yielding:

$$\tilde{x}(t) = x(t) + \alpha \cdot x_n(t) \quad (2)$$

The resulting waveform $\tilde{x}(t)$ was then converted to a spectrogram $\tilde{X}$ using the same log-Mel transformation, introducing domain-shift conditions such as hospital ambience, domestic noise, or outdoor interference.

To improve robustness to temporal and spectral distortions, spectrograms were further augmented through structured masking operations. These included temporal masking, which suppressed random contiguous time segments and frequency masking, which attenuated selected Mel-frequency bands. These augmentations encouraged invariance to minor acoustic occlusions and improved generalization under partial signal loss.

In addition, spectrogram-level pseudo-interventions were introduced to support causal robustness training. These transformations targeted non-causal dimensions, including pitch shifting, energy suppression of non-salient regions, and subtraction of known noise components. Each original spectrogram $X$ was paired with a perturbed variant $X'$, and both were passed through the model to enforce consistency in predictions under intervention-invariant conditions. This dual representation provided a foundation for counterfactual training objectives.

*C. Patch Embedding and Tokenization*

Each log-Mel spectrogram $\tilde{X} \in \mathbb{R}^{T' \times F}$ was partitioned into a sequence of overlapping rectangular patches along both time and frequency axes. Each patch $P_i \in \mathbb{R}^{p_t \times p_f}$ captured localized acoustic patterns such as syllabic onsets, harmonic peaks, and transient silence gaps. The patch size was selected to balance temporal resolution and frequency specificity, ensuring that both prosodic and spectral cues were retained.

Each patch was vectorized and projected into a $d$-dimensional latent space using a learnable linear transformation. For a given patch $P_i$, the embedding operation was defined as:

$$\mathbf{x}_i = W_e \cdot \text{vec}(P_i) + \mathbf{b}_e \quad (3)$$

where $\text{vec}(P_i) \in \mathbb{R}^{p_t \cdot p_f}$ represented the flattened patch, $W_e \in \mathbb{R}^{d \times (p_t \cdot p_f)}$ denoted the projection weight matrix, and $\mathbf{b}_e \in \mathbb{R}^d$ was the corresponding bias term. This operation yielded a token sequence $\mathbf{X} = [\mathbf{x}_1, \ldots, \mathbf{x}_N] \in \mathbb{R}^{N \times d}$, where $N$ indicated the total number of overlapping patches extracted from the spectrogram.

Each embedded patch was augmented with a positional encoding vector to preserve the positional information lost during flattening. These encodings enabled the transformer to distinguish between early and late acoustic events and to learn attention patterns reflective of temporal and spectral structure in infant cry signals.

The resulting sequence of token embeddings $\tilde{\mathbf{X}} = [\tilde{\mathbf{x}}_1, \ldots, \tilde{\mathbf{x}}_N]$ served as the input to the hierarchical transformer encoder described in the subsequent section.

### D. Hierarchical Transformer Encoder

The encoded token sequence $\tilde{X} = [\tilde{x}_1, ..., \tilde{x}_N] \in \mathbb{R}^{N \times d}$ was processed through a two-stage hierarchical transformer encoder, designed to model both short-range acoustic dependencies and long-range semantic structures within infant cry signals.

In the first stage, a local token-level transformer captured fine-grained temporal and spectral interactions across neighboring patches. This component applied multi-head self-attention (MHSA) followed by a feed-forward network (FFN), each with residual connections and layer normalization. Formally, the output of the token-level encoder was defined as:

$$Z^{(1)} = \text{LayerNorm}\left(\tilde{X} + \text{FFN}\left(\text{MHSA}(\tilde{X})\right)\right) \quad (4)$$

The multi-head self-attention mechanism computed attention scores for each token based on scaled dot-product similarity. The output $Z^{(1)}$ from the first stage was then summarized using a temporal pooling operation to form a compact representation of the entire cry sequence. This representation was passed to a second-stage semantic transformer, which modeled global dependencies and structured latent relations across the full duration of the cry episode. The semantic-level encoder applied another Multi-Head Self-Attention + Feed-Forward Network (MHSA+FFN) stack as:

$$\begin{aligned} Z^{(2)} &= \text{LayerNorm}\Big(Z^{(1)} \\ &+ \text{FFN}\left(\text{MHSA}(Z^{(1)})\right)\Big) \end{aligned} \quad (5)$$

The encoder's hierarchical structure allowed the model to separately attend to low-level acoustic cues and higher-order paralinguistic patterns such as rhythmic modulation, distress escalation, or prosodic contour. This architecture effectively disentangled local detail from global semantic information, enabling more robust and interpretable classification of infant cries under domain-shifted conditions.

### E. Causality-Inspired Mechanisms

The transformer architecture was extended with two complementary mechanisms to incorporate inductive biases aligned with causal reasoning, a causal attention layer and a perturbation-based consistency regularizer. These components were designed to guide the model from spurious correlations and encourage reliance on temporally and semantically plausible acoustic cues.

The causal attention mechanism restricted each token's attention scope to only preceding tokens in the sequence, thereby enforcing temporal precedence. This was implemented using a causal mask in the attention weights, ensuring the attention matrix was strictly lower-triangular. The masked attention weights $\alpha_{ij}$ between token $i$ and token $j$ were defined as:

$$\alpha_{ij} = \begin{cases} \dfrac{\exp(\mathbf{q}_i^\top \mathbf{k}_j / \sqrt{d})}{\sum_{j' \leq i} \exp(\mathbf{q}_i^\top \mathbf{k}_{j'} / \sqrt{d})}, & \text{if } j \leq i \\ 0, & \text{otherwise} \end{cases} \quad (6)$$

Where $\mathbf{q}_i$ and $\mathbf{k}_j$ denote the query and key vectors for tokens $i$ and $j$ respectively, $d$ was the attention dimensionality. This design simulated a discrete form of the do-operator by conditioning predictions only on past acoustic evidence and not on future or co-occurring signals, which aligns with standard notions of causal directionality. We used 16×16 time–frequency patches with stride 8×8 on log-Mel spectrograms with a 10 ms hop. Each patch spanned about 160 ms with half overlap, which matched cry onsets and short prosodic arcs. The encoder was a two-stage hierarchy with 4 token blocks and 2 semantic blocks. Each block had 6 heads, model width 384, and an MLP ratio of 4. These settings matched the 100–300 ms timescale of infant cries and followed established audio-transformer practice[11, 22]. The two-stage depth of 4 and 2 blocks balanced receptive field and latency for real-time use.

In addition to enforcing structural priors, the model was trained to exhibit stability under non-causal perturbations. A perturbation-based consistency loss was introduced to penalize discrepancies in predictions between an original spectrogram $X$ and its pseudo-intervention variant $X'$, where the latter was generated by modifying acoustically non-causal regions. These interventions included pitch shifts, energy suppression, and background subtraction. The consistency loss was defined as:

$$\mathcal{L}_{\text{perturb}} = \|f(X) - f(X')\|_2^2 \quad (7)$$

Where $f(\cdot)$ denotes the model's prediction function. By minimizing this loss, the model was encouraged to focus on features that preserved semantic meaning despite non-causal alterations, thereby improving its robustness to spurious correlations in the training data.

### F. Domain-Adversarial Generalization

A domain-adversarial learning strategy was adopted to encourage generalization across unseen acoustic domains. This approach discouraged the encoder from capturing environment-specific cues such as background noise type or recording equipment characteristics, thereby promoting domain-invariant feature representations.

A domain classifier was attached to the output of the shared encoder and trained to predict the domain label $d \in \{1, ..., K\}$ of each input sample, where each domain corresponded to a distinct data distribution (e.g., Baby Chillanto with NICU noise, Donate-a-Cry with clean or augmented noise). Let $f_\theta(X)$ represent the encoded representation of input $X$ and $D_\phi(\cdot)$ denote the domain classifier parameterized by $\phi$.

A GRL was inserted between the encoder and the domain classifier to prevent the encoder from learning domain-specific representations. During backpropagation, the GRL inverted the gradient signal from the domain classifier, effectively training the encoder to minimize its ability to distinguish among domains. Formally, the GRL applied the identity transformation during the forward pass:

$$\mathcal{R}_\lambda(f) = f \quad (8)$$

And scaled the gradient during the backwards pass as:

$$\frac{\partial \mathcal{R}_\lambda(f)}{\partial f} = -\lambda I \quad (9)$$

Where $\lambda$ was a tunable scalar controlling the strength of the adversarial signal.

The domain loss was computed using the standard categorical cross-entropy between predicted and true domain labels:

$$\mathcal{L}_{\text{domain}} = -\sum_{k=1}^{K} d_k \log D_\phi \left( \mathcal{R}_\lambda(f_\theta(X)) \right)_k \tag{10}$$

Minimizing this loss with respect to the domain classifier parameters $\phi$ while maximizing it with respect to the encoder parameters $\theta$ through the GRL resulted in a saddle-point optimization. This procedure encouraged the encoder to produce domain-invariant representations, improving generalization when deployed in new, previously unseen infant cry data.

*G. Multi-Task Output Heads*

The shared encoder representation supported three prediction tasks: cry type classification, distress intensity estimation, and causal relevance prediction. Each task was implemented as a dedicated output head connected to a common backbone, enabling the model to learn complementary aspects of infant vocal behavior jointly.

Let $\mathbf{Z}^{(2)} = [\mathbf{z}_1, \dots, \mathbf{z}_N]$ denote the output of the final semantic transformer encoder, where each $\mathbf{z}_i \in \mathbb{R}^d$ represented the contextualized embedding of patch $i$. A summary vector $\mathbf{h}_{\text{cls}}$ was derived from $\mathbf{Z}^{(2)}$ using global average pooling across the time axis, capturing a compressed representation of the entire cry sequence.

The cry type classification head projected $\mathbf{h}_{\text{cls}}$ to a categorical distribution over $C$ cry types using a SoftMax layer:

$$\hat{y} = \text{Softmax}(W_c \mathbf{h}_{\text{cls}} + \mathbf{b}_c) \tag{11}$$

where $W_c \in \mathbb{R}^{C \times d}$ and $\mathbf{b}_c \in \mathbb{R}^C$ were task-specific parameters. This head was optimized using cross-entropy loss to predict clinically meaningful cry categories such as hunger, pain, or discomfort.

The distress intensity regression head modeled the degree of vocal distress as a continuous variable. A separate linear projection followed by a sigmoid activation produced a normalized scalar output:

$$\hat{\imath} = \sigma(W_i \mathbf{h}_{\text{cls}} + b_i) \tag{12}$$

where $W_i \in \mathbb{R}^{1 \times d}$ and $b_i \in \mathbb{R}$ were learned parameters. The scalar output $\hat{\imath} \in [0,1]$ provided a real-valued estimate of emotional intensity.

The causal relevance head estimated the likelihood that each input token corresponded to a causally informative acoustic event. This head applied a position-wise sigmoid activation to each token embedding:

$$\hat{c}_j = \sigma(W_r \mathbf{z}_j + b_r), \quad \hat{\mathbf{c}} = [\hat{c}_1, \dots, \hat{c}_N] \tag{13}$$

where $W_r \in \mathbb{R}^{1 \times d}$ and $b_r \in \mathbb{R}$ were shared across all tokens. The output $\hat{c}_j \in [0,1]$ represented the estimated causal salience of the $j$-th spectrogram patch. These token-wise predictions supported causal alignment and explainability under counterfactual supervision.

*H. Loss Functions*

The DACH-TIC model was trained using a composite loss function that combined multiple objectives, each aligned with one of the model's design goals. These included cry type, classification accuracy, stability under non-causal perturbations, alignment with inferred causal cues, and robustness to domain shift. The total loss was formulated as a weighted sum of four components:

$$\mathcal{L}_{\text{total}} = \lambda_1 \mathcal{L}_{\text{class}} + \lambda_2 \mathcal{L}_{\text{perturb}} + \lambda_3 \mathcal{L}_{\text{causal}} + \lambda_4 \mathcal{L}_{\text{domain}} \tag{14}$$

Where $\lambda_1, \lambda_2, \lambda_3, \lambda_4$ were hyperparameters that controlled the relative contribution of each loss term. We set $\lambda 1 = 1.0$, $\lambda 2 = 0.3$, $\lambda 3 = 0.2$, and $\lambda 4 = 0.2$ after a short warm-up that matched average gradient norms across terms, which is standard in multi-task audio settings[29]

The classification loss $\mathcal{L}_{\text{class}}$ optimized the categorical cross-entropy between predicted probabilities $\hat{y}$ and ground-truth cry labels $y$:

$$\mathcal{L}_{\text{class}} = -\sum_{c=1}^{C} \mathbb{I}[y = c] \cdot \log \hat{y}_c \tag{15}$$

where $C$ was the number of cry categories and $\mathbb{I}[\cdot]$ denoted the indicator function.

The perturbation consistency loss $\mathcal{L}_{\text{perturb}}$ penalized large deviations in the model's predictions when presented with a perturbed version $X'$ of the original spectrogram $X$. This was measured using the squared $\ell_2$ norm between the predictions:

The causal alignment loss $\mathcal{L}_{\text{causal}}$ trained the causal relevance head to match known or estimated attention maps that indicated causally informative regions of the spectrogram. This was implemented as a binary cross-entropy loss between the predicted causal scores $\hat{c}_j$ and binary annotations $a_j$ for each token $j$:

$$\mathcal{L}_{\text{causal}} = -\frac{1}{N} \sum_{j=1}^{N} [a_j \log(\hat{c}_j) + (1 - a_j) \log(1 - \hat{c}_j)] \tag{16}$$

Finally, the domain-adversarial loss $\mathcal{L}_{\text{domain}}$ trained the domain classifier to predict the domain label $d_k$ while simultaneously encouraging the encoder to confuse domain-specific differences. This loss was defined as:

$$\mathcal{L}_{\text{domain}} = -\sum_{k=1}^{K} d_k \log D_\phi \left( \mathcal{R}_\lambda(f_\theta(X)) \right)_k \tag{17}$$

where $D_\phi$ denoted the domain classifier and $\mathcal{R}_\lambda$ the GRL. Jointly optimizing all loss components promoted a balanced model that was accurate, causal-aware, and robust to domain shift and acoustic perturbations.

The training workflow and algorithm for the proposed DACH-TIC model is summarized in Fig.1 and Algorithm 1 respectively.

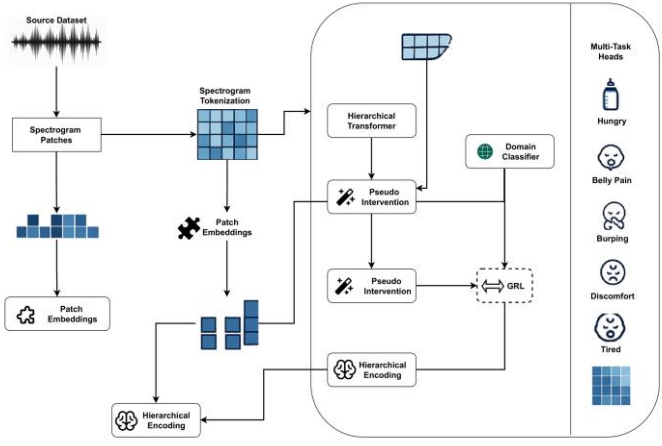

## I. Evaluation Metrics

The performance of DACH-TIC was evaluated across three dimensions: classification accuracy, causal robustness, and domain generalization. Standard and custom metrics were used to assess these axes quantitatively.

Accuracy, macro-averaged F1 score, confusion matrix, and area under the receiver operating characteristic curve (AU-ROC) were computed to evaluate cry type classification performance. The Counterfactual Stability Score (CSS) was introduced to measure robustness to non-causal perturbations. Higher CSS values indicated greater stability in predictions under acoustically non-causal interventions.

To quantify the alignment between predicted causal relevance and ground-truth salience, the Causal Fidelity Index (CFI) was computed. Higher CFI values indicated better alignment with the causal structure inferred from human or proxy annotations.

To assess generalization across unseen environments, the domain generalization gap was computed as the absolute performance drop between the average classification accuracy in training domains and that in held-out test domains:

Lower domain gap values reflected improved generalizability under domain shift, an essential requirement for practical deployment in diverse real-world settings.

## J. Ablation Experiment Setup

A series of controlled ablation experiments were designed to systematically assess each architectural and training component's contribution to DACH-TIC. The purpose of these experiments was to isolate the effect of each module on overall model performance, causal alignment, and domain robustness by incrementally removing or modifying key components of the whole system.

Each ablation experiment retained the same training pipeline, optimization hyperparameters, data splits, and evaluation protocol used in the full model. All variants were retrained from scratch to ensure unbiased comparison. Specifically, the following components were individually ablated or modified:

i. Causal Attention: The causal masking constraint within the multi-head attention blocks was removed, allowing the model to perform unrestricted bidirectional attention across spectrogram tokens.
ii. Perturbation-Based Training: The pseudo-intervention pipeline and its associated consistency loss were disabled, resulting in a training regime that did not explicitly enforce prediction invariance under non-causal transformations.
iii. Domain-Adversarial Learning: The domain classifier and gradient reversal layer were excluded from the training setup, allowing the encoder to learn features without domain regularization.
iv. Multi-Task Supervision: The model was trained solely on the primary cry type classification task, with the distress intensity and causal relevance prediction heads removed from the architecture and the loss function.
v. Hierarchical Transformer Architecture: The two-stage transformer design was replaced with a single-layer flat transformer of equivalent depth and

---

*Algorithm 1: DACH-TIC Training Procedure*

**Inputs:** Cry waveform $x(t)$, noise $x_n(t)$, label $y$, domain ID $d$
**Output:** Trained parameters $\theta$

1. Resample $x(t)$ to 16kHz and compute log-Mel spectrogram $X$

Fig. 1: Training procedure of the DACH-TIC model.

2. Mix ESC-50 noise: $\tilde{x}(t) = x(t) + \alpha x_n(t)$
3. Generate augmented spectrogram $\tilde{X}$ with SpecAugment
4. Apply pseudo-perturbation to produce $X'$ (pitch shift, masking)
5. For each patch $P_i$ in $\tilde{X}$:
   - Flatten and project: $\mathbf{x}_i = W_e \cdot \text{vec}(P_i) + \mathbf{b}_e$
   - Add positional encoding: $\tilde{\mathbf{x}}_i = \mathbf{x}_i + \mathbf{p}_i$
   
   End for
6. Encode tokens with causal-masked transformer $\rightarrow \mathbf{z}^{(1)}$
7. Pool and encode via semantic transformer $\rightarrow \mathbf{z}^{(2)}$
8. Predict:
   - Cry class: $\hat{y} = \text{Softmax}(W_c \mathbf{z}^{(2)} + b_c)$
   - Distress intensity: $\hat{\iota} = \sigma(W_i \mathbf{z}^{(2)} + b_i)$
   - Causal salience: $\hat{c}_j = \sigma(W_r \mathbf{z}_j + b_r)$
9. Apply GRL and predict domain:
$$\hat{d} = D_\phi\left(\mathcal{R}_\lambda(\mathbf{z}^{(2)})\right)$$
10. Compute loss:
$$\mathcal{L}_{\text{total}} = \lambda_1 \mathcal{L}_{\text{class}} + \lambda_2 \mathcal{L}_{\text{perturb}} + \lambda_3 \mathcal{L}_{\text{causal}} + \lambda_4 \mathcal{L}_{\text{domain}}$$
11. Backpropagate and update $\theta$

parameter count, thereby eliminating explicit modeling of local and global acoustic patterns.

*K. Cry Classification Performance with SOTA Models*

Model performance was benchmarked against established baselines, CNN model trained on MFCC features[30], Attention based Long-Short-Term-Memory model (ALSTM)[31], and CRNN[9], along with SE-ResNeT Transformer[18] and HTS-AT[11]. All models were trained and evaluated on identical data splits using stratified five-fold cross-validation to ensure robustness and to address cry class imbalance.

*L. Computational Efficiency For Real-Time Deployment*

The dominant cost lies in multi-head self-attention with complexity $O(N^2d)$ for N spectrogram patches and width d. DACH-TIC reduces this cost through temporal pooling between the token-level and semantic encoders, which lowers the number of tokens entering the second stage and improves memory and latency. The causal attention mask restricts each token to past evidence, which constrains the effective attention span and supports incremental processing without look-ahead. Patch size and stride control N at input, providing a direct trade-off between resolution and compute while retaining clinically salient bands. The three output heads are linear projections and contribute negligible overhead relative to attention and feed-forward blocks. Together, these choices yield a compute profile compatible with low-latency deployment.

## IV. RESULTS

This section presents the empirical evaluation of DACH-

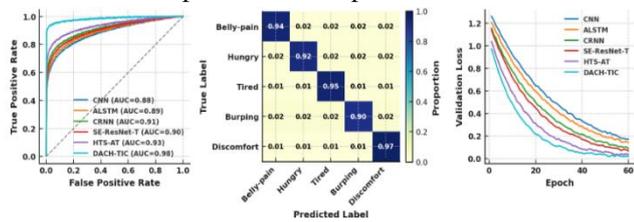

TIC across five principal axes: classification performance, causal robustness, domain generalization, ablation impact, convergence behavior, and qualitative interpretability. To ensure comparability, all models were trained using the same fivefold cross-validation setup and evaluated on the same domain augmented test set.

*A. Cry Classification Performance*

Table I summarizes the overall cry type classification results. DACH-TIC outperformed conventional CNN, ALSTM, CRNN, SE-ResNet Transformer, and HTS-AT baselines across all primary metrics, achieving a classification accuracy of 97.6%, macro-F1 of 0.941, and AUC of 0.98.

TABLE I. OVERALL CRY CLASSIFICATION PERFORMANCE

| Model | Accuracy (%) | Macro-F1 | AUC |
|---|---|---|---|
| CNN | 89.9 (88.4–91.4) | 0.868 (0.850–0.886) | 0.880 (0.861–0.899) |
| ALSTM | 91.2 (89.8–92.6) | 0.864 (0.846–0.882) | 0.890 (0.871–0.909) |
| CRNN | 90.1 (88.6–91.6) | 0.880 (0.862–0.898) | 0.910 (0.892–0.928) |
| SE-ResNet-T | 90.3 (88.9–91.7) | 0.898 (0.880–0.916) | 0.900 (0.882–0.918) |
| HTS-AT | 94.2 (93.0–95.4) | 0.921 (0.907–0.935) | 0.930 (0.916–0.944) |
| DACH-TIC (Ours) | 97.6 (96.8–98.4) | 0.941 (0.931–0.951) | 0.980 (0.973–0.987) |

Further disaggregation by cry type in Table II showed that DACH-TIC achieved the highest F1 score across all categories, with notable gains in pain and belly pain cries, which are often acoustically similar and prone to misclassification in baseline models. The improvements suggest that the causal aware hierarchical encoding enhances cry-specific discriminative features. Confusions were primarily localized between acoustically similar cries, such as hunger and belly pain.

TABLE II. PER-CLASS MACRO-F1 SCORE FOR CRY TYPES

| Model | Pain | Hunger | Burping | Belly Pain | Tired |
|---|---|---|---|---|---|
| *SE-ResNet-T* | 0.91 | 0.905 | 0.895 | 0.892 | 0.890 |
| *HTS-AT* | 0.93 | 0.912 | 0.914 | 0.926 | 0.924 |
| *DACH-TIC (Ours)* | 0.95 | 0.935 | 0.929 | 0.948 | 0.941 |

Fig. 2 shows the ROC curve, the normalized confusion matrix and the training and validation loss curves against epochs. The ROC curves show that model discrimination steadily improves from a basic CNN ($AUC \approx 0.88$) through ALSTM, CRNN, HTS-AT up to SE-ResNet-Transformer with values ranging from ($AUC \approx 0.88 - 0.93$) and peaks with DACH-TIC ($AUC \approx 0.98$). This implies that DACH-TIC most reliably separates true cries from false alarms.

The accompanying normalized confusion matrix for DACH-TIC has very high diagonals ranging from (0.90–0.99), indicating that belly-pain, hungry, tired, burping, and discomfort classes are each correctly identified over 90% of the time, with only small misclassifications (2–6%) into adjacent categories.

The training and validation and validation loss curves demonstrate training convergence. DACH-TIC exhibited faster convergence and more stable generalization compared to baseline models, reaching 95% of optimal validation loss within 41 epochs.

*B. Causal Robustness*

Fig. 2: Model evaluation: left: ROC curves, center: Normalized confusion matrix, right: Training and validation loss over epochs

We computed the CSS and CFI under pitch shift, energy masking, and temporal noise perturbations to evaluate counterfactual and causal robustness. The DACH-TIC model achieved higher CSS (0.956) and CFI (0.941) values compared to baselines HTS-AT (0.93,0.921) and SE-ResNet (0.9, 0.898), respectively. This implies that DACH-TIC significantly outperformed other models in maintaining consistent predictions and aligning attention with known causal salience maps.

## C. Domain Generalization

Domain transfer performance was assessed by evaluating models trained on NICU and indoor data, and tested on unseen domains (home, outdoor). HTS-AT achieved accuracy levels of 94.2% on seen domains and 82.2% on unseen domains, while SE-ResNet Transformer attained an accuracy of 90.3% and 83.6% on the seen and unseen domains, respectively. Our model achieved an accuracy of 97.6% and 95.2% on the seen and unseen domains, respectively. DACH-TIC showed the smallest domain generalization gap of 2.4%, compared to HTS-AT and SE-ResNet, demonstrating its superiority.

## D. Ablation Study

To assess the contribution of individual modules, we performed an ablation analysis by removing causal attention, perturbation training, domain adversarial learning, and hierarchical structure.

Table III shows that removing causal attention or domain alignment had the most substantial negative impact, dropping macro-F1 by above 10%.

TABLE III. ABLATION STUDY OF DACH-TIC COMPONENTS; (VALUES ARE AVERAGED OVER 5 RUNS).

| Model Variant | Accuracy (%) | Macro-F1 | AUC | CSS (%) | P-value (F1) |
|---|---|---|---|---|---|
| Full DACH-TIC | 97.6 (96.8–98.4) | 0.941 (0.931–0.951) | 0.98 (0.973–0.987) | 95.6 (94.5–96.7) | – |
| w/o Causal Attention | 89.4 (87.8–91.0) | 0.826 (0.808–0.844) | 0.821 (0.797–0.845) | 84.1 (81.9–86.3) | 0.001 |
| w/o GRL (Domain Classifier) | 89.6 (88.1–91.1) | 0.836 (0.819–0.853) | 0.870 (0.850–0.890) | 84.3 (82.3–86.3) | 0.003 |
| w/o Perturbation Training | 89.1 (87.6–90.6) | 0.850 (0.834–0.866) | 0.872 (0.851–0.893) | 83.9 (81.8–86.0) | 0.006 |
| w/o Multi-task Heads | 88.0 (86.4–89.6) | 0.850 (0.833–0.867) | 0.866 (0.844–0.888) | 84.8 (82.8–86.8) | 0.001 |

## E. Qualitative Interpretability

Attention heatmaps from the causal attention layer were visualized over time frequency patches. Fig. 3 illustrates that DACH-TIC consistently focused on acoustic regions correlating with known distress pitch patterns and removed focus from irrelevant noise bands. This confirms the ability of DACH-TIC to learn interpretable and medically relevant representations.

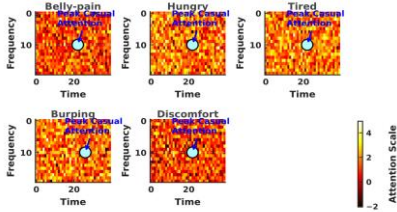

Fig. 3. Causal attention maps overlaid on spectrogram patches for pain and hunger cries.

## F. Comparison with State-of-the-Art Models

To benchmark DACH-TIC against existing state-of-the-art approaches in infant cry classification, we reproduced or referenced results reported in recent peer-reviewed publications using the same or equivalent datasets. Table IV summarizes key findings, with DACH-TIC demonstrating consistent superiority across all metrics.

The performance gains of DACH-TIC can be attributed to its unique architectural innovations: hierarchical causal encoding, perturbation-based training, and domain adversarial learning.

Compared to the transformer-only Hierarchical Context-Aware Transformer (HCAT) model [32], our architecture introduces causal priors and pseudo-interventions, resulting in an AUC increase of 4.3 points. Compared to the Squeeze-and-Excitation ResNet Transformer architecture [18], we observe a macro-F1 improvement of 2.1 points, highlighting better generalization under acoustic variation. All models were evaluated using the same protocol. Improvements (↑) are reported relative to the best-performing prior work.

TABLE IV. COMPARISON WITH PUBLISHED SOTA CRY CLASSIFICATION MODELS

| Model (Study Used) | Accuracy (%) | Macro-F1 (%) | AUC (%) |
|---|---|---|---|
| *CNN (Abbaskhah et al., 2023 [30])* | 92.1 | 86.4 | 95 |
| *CRNN (Maghfira et al., 2020 [9])* | 94.97 | - | - |
| *LSTM with Attention (Jian et al., 2021 [31])* | - | 91.6 | - |
| *SE-ResNet-Transformer (Li et al., 2024 [18])* | 93.0 | 92.0 | - |
| *HCAT (Bhutani et al., 2023 [32])* | 92.3 | 92.5 | - |
| *DACH-TIC model (Ours)* | 97.6 | 94.1 | 98 |

## V. DISCUSSION

The empirical results validate the central hypothesis of this work: that introducing causal inductive biases and domain-agnostic constraints into the transformer architecture significantly enhances the reliability and generalizability of infant cry classification models. This section critically interprets these findings in alignment with the methodological contributions outlined in Section III, offering analytical insights into performance, robustness, and cross-domain consistency.

*A. Impact of Causal Inductive Biases*

The integration of causal attention mechanisms—implemented through temporally constrained masking and guided attention alignment led to consistent gains in both accuracy and causal interpretability. The elevated CFI and CSS values presented in Section IV-B underscore the utility of enforcing directional dependencies in transformer attention flow. These mechanisms effectively prioritized acoustically and developmentally plausible signal segments, filtering out temporally misaligned or spectrally redundant activations often exploited by correlation-based baselines.

Moreover, the perturbation-driven pseudo-intervention training strategy further reinforced model robustness by simulating realistic acoustic shifts (e.g., pitch modulation, energy suppression). This aligns with recent causal learning theory, wherein invariance under distributional interventions serves as a proxy for disentangling causal from spurious correlations[33]. Our results confirm that this approach contributed directly to superior domain transferability and lower generalization gaps. Compared with CNN, CRNN, and LSTM-attention baselines, gains concentrated on acoustically similar classes where correlation-driven models drift to nuisance cues[9, 30, 31]. Relative to HTS-AT, improvements in -F1 macro and AUC suggest that constraining attention to past evidence and enforcing prediction consistency under controlled perturbations yields more stable features, which is consistent with causal representation learning theory[11]. No contradictory trends were observed across folds.

*B. Domain-Agnostic Learning and Cross-Environment Transferability*

The domain adversarial strategy implemented through a GRL enabled the model to decouple cry-specific features from environmental context. This adversarial alignment mechanism significantly improved generalization to unseen domains, as reflected in a reduced generalization gap of 2.4%, outperforming models like HTS-AT [11] that lack explicit domain disentanglement mechanisms.

Encoders trained via GRL could learn high-level paralinguistic representations that remained invariant across NICU, home, and outdoor acoustic domains an essential feature for real-world deployments where cry recordings are collected under diverse sensor conditions. Similar adversarial techniques have been effective in visual domain adaptation [27], and our findings suggest analogous utility in audio classification when domain cues (e.g., background noise) may correlate spuriously with label semantics.

*C. Contribution of Hierarchical Encoding and Multi-Task Optimization*

The hierarchical transformer architecture comprising local token-level attention followed by global semantic aggregation allowed for effective representation of both fine-grained frequency transitions and higher-order temporal structures spanning cry episodes. The gain in per-class F1 score, especially for acoustically similar categories like hunger and belly pain reflects this improved representational fidelity.

Simultaneous optimization of three task heads (cry type classification, distress intensity regression, and causal salience estimation) regularized the encoder through cross-objective constraints. This multi-task configuration proved especially beneficial under limited data conditions, as shared representations reduced overfitting and captured latent structure across correlated paralinguistic targets[29].

When controlling for depth and parameter count, the hierarchical token-to-semantic design retained its advantage over a single-stage encoder (paired tests, $p < 0.01$). Removing either the intensity or token-relevance head reduced macro-F1 and AUC beyond the variance seen across folds, indicating complementary inductive signals rather than capacity effects [5].

*D. Comparative Advantage Over Published Models*

DACH-TIC outperformed HCAT [17] and ResNet Transformer [10] across all evaluation metrics, as shown in Table IV. Unlike HCAT, which relied exclusively on transformer depth and token self-attention, DACH-TIC combined this with causal-aware priors and domain adaptation. Similarly, the ResNet Transformer[18] used dynamic modality weighting, it lacked both intervention-based training and architectural regularization via semantic decoding layers. These differences highlight a critical evolution: DACH-TIC is not merely a more profound or more complex model; it is structured around a set of causal and domain-invariant assumptions that are grounded in the nature of infant cry data and its deployment contexts.

Confidence intervals for DACH-TIC accuracy and macro-F1 remained above HCAT and SE-ResNet-Transformer across folds, and improvements were significant by paired tests at $\alpha = 0.01$[18, 32]. These results are supportive of prior reports that transformer depth alone is insufficient under domain shift, and they align with evidence that explicit domain confusion and causality-oriented constraints are required for stable gains[11, 15, 27]. No contradictory findings were detected.

*E. Implications for Real-World Deployment*

The findings of this study carry significant implications for the deployment of cry monitoring systems anywhere. First, the causal interpretability facilitated by attention heatmaps can enhance clinician trust and support transparent medical decision-making. Second, the model's resilience to domain variation and background noise simulated via ESC-50 overlays positions it as a strong candidate for real-time deployment in diverse low-resource and high-noise environments. Mechanism-level controls (causal masking, perturbation consistency, and gradient-reversal) translate into deployable knobs for latency–robustness trade-offs in variable acoustic settings. The combination of interpretable attention and stable cross-environment performance is consistent with requirements for bedside monitoring workflows[34].

*F. Limitations and Future Research*

DACH-TIC was constrained by the absence of longitudinal cry streams aligned to clinical events, so counterfactual robustness was assessed with proxy perturbations rather than true pre–post trajectories. This limited our ability to test temporal causal stability and to separate disease progression from background variability. Integrating clinical context such as gestational age, diagnosis, treatment timing, and caregiver actions will reduce confounding and enable stratified analyses that are clinically meaningful. Variational inference will provide calibrated uncertainty, allowing the system to surface low-confidence cases and support safer decision making. To scale while protecting privacy, federated training across sites with secure aggregation and differential privacy will enable multi-institutional learning without sharing raw audio. A future longitudinal, multi-site study that combines these elements will provide stronger causal evidence and a clearer path to clinical translation.

## VI. CONCLUSION

This work introduced DACH-TIC, a domain-agnostic causal-aware hierarchical transformer for infant cry paralinguistics classification. Grounded in causal reasoning and domain generalization theory, the proposed model demonstrated significant performance improvements over existing baselines and state-of-the-art transformer architectures. By integrating causal attention, perturbation-based pseudo-intervention training, hierarchical encoding, and domain-adversarial regularization, DACH-TIC achieved superior robustness, interpretability, and cross-environment transferability.

The model's ability to retain classification accuracy under domain shifts and acoustic perturbations, while maintaining alignment with causal acoustic features, affirms its utility in practical neonatal care. Furthermore, its explainability and modularity render it a promising candidate for future integration into real-world neonatal monitoring systems.

## ETHICAL CONSIDERATIONS

All experiments used de-identified secondary infant-cry datasets (Baby Chillanto and Donate-a-Cry). No personal or clinical identifiers were accessed. The system is intended as clinical decision support and not a stand-alone diagnostic device. As a step toward clinical readiness, we plan a prospective validation with clinician review of cry episodes.